\documentclass[12pt]{article}
\usepackage{graphicx}
\usepackage{tikz}
\usepackage{ifpdf}
\begin{document}
\begin{center}
{\Large Holography constrains quantum bounce}\\
\bigskip
{\bf D.H. Coule}\\
\bigskip
Institute of Cosmology and Gravitation,\\Dennis Sciama Building,
Burnaby Rd,  University of Portsmouth, Portsmouth  PO1 3FX, UK\\
\bigskip

\begin{abstract}

Recent work  in quantum loop cosmology suggests the universe
undergoes a bounce when evolving from a previously collapsing
phase. However, with matter sources that obey the strong-energy
condition i.e. non-inflationary, the scenario appears to strongly
violate the holographic bound on entropy $S\leq A/4$ during the
bouncing phase, where $A$ now represents the cross-sectional area
of the bounce.

 We also give a simple argument why any inflationary
phase after the bounce is unlikely due to the prior dissipation of
any kinetic dominated scalar field with large frequency $m\sim
10^{-5}$  over semi-infinite collapsing time scales. Rather
entropy increase should be expected and any  subsequent
inflationary phase would have entailed a violation of the
generalized second law before the bounce occurs.


\end{abstract}
\end{center}
\newpage

In some  recent approaches to quantum gravity the usual
singularity in big bang cosmology is possibly replaced by a bounce
due to the influence of  quantum gravitational effects. As an
approximation to this phenomena the standard Friedmann equation
for a FRW model e.g.[1,2]

\begin{equation}
H^2+\frac{k}{a^2} =\rho
\end{equation}
can become   modified such that
\begin{equation}
H^2+\frac{k}{a^2} =\rho-\frac{\rho^2}{\rho_c}
\end{equation}
where $\rho_c$ represents the critical energy scale, typically
expected to be around the  Planck scale.\footnote{ We use Planck
units throughout and set numeric factors like $8\pi/3$ to unity.}
This behaviour occurs
 in  loop quantum gravity [3-6] although in that case the curvature dependence
 is actually slightly more involved, but for our purposes this
 simplified equation will first  prove adequate.

  A similar  description
  might  be obtained with  brane models with an extra time
dimension [7] although this is probably observationally discounted
[8]. Note that a suggested [9] single negative tension brane  is
not suitable: it differs from eq.(2) by an overall minus sign on
the R.H.S. since starting with a positive 5-dimensional Planck
mass the negative tension causes the 4-dimensional Newton's
constant to become negative cf.[10].

The holography bound [11] on entropy $S\leq A/4$, where $A$ is the
corresponding area measured in Planck units  appears saturated for
black holes but with usual matter there is a stronger restriction
$S\leq A^{3/4}$ - see e.g.[12] for extensive reviews. By choosing
a closed radiation dominated FRW universe with corresponding
maximum size $a_{max} \sim A^{1/2}$ this stronger bound is just
saturated and $a^2_{max} \sim S_{r}^{4/3}$, where $S_{r}$
represents the corresponding entropy of the radiation present,
ignoring some numerical factors that correspond to the number of
spin states cf.[13-15].

By starting at the maximum size of a closed radiation dominated
universe we then  know the amount of entropy initially present, By
letting the model collapse we can find the corresponding minimum
radius $a_{min}$ and the corresponding allowed entropy at the
bounce $S_{b}$ . Since we know the amount of entropy, and assuming
adiabatic behaviour so that the entropy remains constant, we can
check whether the holography bound $S\leq A/4$ remains satisfied
or not during all the ensuing evolution. Of course, with the
standard Friedmann equation  the bound will become violated but
this simply represents the impending  the big crunch singularity
which should now  be evaded by the modifications to the equations.

As a first approximation we first keep the  usual curvature  term
$k/a^2$ on the LHS of eq.(2). Also since the matter behaviour is
expected to display its usual form we take $\rho=\alpha/a^4$. Then
$a_{max}^2\sim \alpha$ and assuming $\rho_c\simeq 1$, we find
$S_{b}\simeq  a_{min}^2 \sim \alpha^{1/2}$. Then since $\alpha
^{3/4} >\alpha^{1/2}$ the holography bound is strongly violated
i.e. $S_{r}>S_{b}$ . We typically have in mind starting with a
large classical universe with $\alpha\sim 10^{120}$, so the
starting entropy of the radiation is $10^{90}$ but only a value of
$10^{60}$ should be possible if the holographic bound is still
valid across the constricted bounce. This result apparently
contradicts those obtained previously in ref.[28]. But there they
erroneously claim that $\alpha$ (their $K_o$ constant of eq.(2.4),
(3.8) and (3.10)  is some ``integration constant without physical
significance'' but rather this constant determines the maximum
size and hence entropy content of a closed FRW model.

To correctly include  curvature $k=1$ in loop quantum cosmology
actually involves a more complicated equation so that eq.(2)
becomes [5,16]
\begin{equation}
H^2=\left ( \rho-\frac{1}{a^2}\right ) \left(
1-\frac{\rho}{\rho_c}+\frac{1}{\rho_c a^2}\right )
\end{equation}

For the case of a massless scalar field the corresponding values
are $a_{max}^2 \sim p_{\phi}$ and $a_{min}^2 \sim p_{\phi}^{2/3}$
[5]. The parameter $p_{\phi}$ is here analogous to the previous
$\alpha$. The corresponding holography bound is now violated, but
less severely, since now the corresponding values of entropy are
$p_{\phi}^{3/4}
> p_{\phi}^{2/3}$.

 A more careful analysis of eq.(3), together with  other relevant equations presented in ref.[5,16], confirms
  the results and shows that the
 apparent violation becomes larger as the strong-energy condition boundary is
 approached and that the bound could become satisfied for
 ultra-stiff equations of state i.e $w>4/3$, where  for a  perfect fluid $p=w\rho$. Such equations of
 state have previously been used during the collapsing stage of
 the Ekpyrotic scenario [17] with certain possible advantages [18].

   The use of a closed model was not strictly relevant for
 this argument since one could start with any finite physical  size
  and derive analogous quantities. The crucial
 point is that the universe is being constricted into a finite
 size during the bounce regardless of the underlying  geometry
 $k$. This in turn simplifies the  application of  holography bound
 compared to various complications that can occur in usual FRW models [12].

\begin{figure}
\includegraphics[width=6cm]{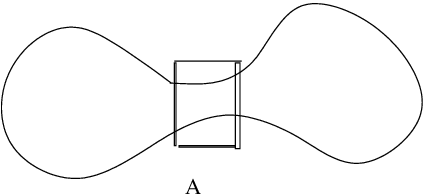}
\caption{  Sketch of the universe ``threading the needle'':
bouncing at a size below the holographic bound of area   A}
\end{figure}

 The bound might also be satisfied if the critical density $\rho_c$
  is vastly decreased, so the repulsive gravitational effect can intervene sooner,
   but this is probably anyway inconsistent with other phenomena.
   There might also be scope to increase the bounce size $a_{min}$
   by employing different quantization procedures that introduce
   further  dependence on the  lattice size employed. Although so-called polymer
  quantization  appears to then  actually reduce the bounce size
  [19]. It is worrying, though, that the procedures have already meant to
  have agreed on the standard Bekenstein-Hawking black hole result $S=A/4$  when setting  the
  arbitrary Barbero-Immirzi parameter, which in term gives the corresponding  $\rho_{c}$ value,  see e.g.[20].

 If black holes are also present, and during the collapsing phase they
 might be expected to more easily form and  congeal cf.[15], the bound will be much
 harder to satisfy since the entropy
 can potentially increase towards  $S\sim a_{max}^2 \sim \alpha \sim 10^{120}$
 : essentially the matter during the bounce is being  more compressed  than
 in  the usual black hole case, such densely entropic
  states have previously been dubbed monsters [21]. One can see this potential problem by simply
   relating the mass of the observable
  universe $\sim 10^{55}kg$ to the Planck density $\sim 10^{97}kgm^{-3}$ so the universe
  could be apparently squashed into a volume
  $10^{-42} m^3$ if one could work with Planck density matter: yet the Schwarzschild radius of a supermassive black hole
  is alone $\sim 10^{10}m$ so we never remotely deal with Planck energy densities with large black holes: the average density
  of a large black hole can be small!

    Another way
 to evade this entropy problem ( and this might be the cause of the discrepancy with ref.[28] ) would be to have only Planck sized
 quantities during the collapsing phase i.e. $ \alpha\sim p_{\phi} \sim
 1$ but then the universe, bereft of classical matter before the bounce,  is vastly asymmetric cf.[22].
 An inflationary phase would have to ensure sufficient entropy
 production for the later universe.

  The question
 of how much information still remains in the post-bounce phase  from the previous collapsing stage will
 obviously be strongly dependent on preserving entropy across the
 bounce - in the example with radiation only a fraction
 $10^{60}/10^{90}\sim 10^{-30}$ is preserved if the bound is to be
 satisfied - so only a limited knowledge of the collapsing phase would still be possible cf.[23].

  Whether this ``eye of the needle'' constriction -see Fig.(1)- is too severe and the holography bound
  should instead be respected; or else other
  processes intervene to reduce the entropy and so violate the generalized second law, will have to be
  resolved before such  a  (monster?) quantum bounce can be countenanced.

  Let us consider further why the possibility of inflation after
  the bounce is difficult to envision. The model is envisioned to display the
  following stages of first scalar field `fast-roll' kinetic, or strictly speaking the oscillatory behaviour around the potential's minimum (Osc) with effectively a $p=0$ equation of state [1,2], then potential $V(\phi)$ domination  and finally matter production, where matter represents the eventual
  production of standard matter components. The model is therefore claimed to have the following stages of development\\
  \\
  Osc $\rightarrow$ Potential $\rightarrow$ Osc
  $\rightarrow$ Matter\\
  \\

  Firstly there is an
  argument using the canonical measure [24] that suggests inflation, driven for example
  by a massive scalar field $V(\phi)=1/2m^2\phi^2$,  is
  likely if applied at the bounce point [25]. It is known that applying
  such an argument at Planck energy densities is generally
  conducive to obtaining inflationary behaviour [26]. But this argument
   is akin to the universe suddenly commencing its
  existence at the bounce itself. The canonical measure is only valid for non-dissipative Hamiltonian systems so it can only be applied in certain situations.  
  The analysis should rather take
  into account the previous mostly classical evolution of the universe collapsing
  towards the bounce having started presumably at time
  $t=-\infty$. Incidentally in [24] they divide out by a claimed irrelevant ``gauge''  $a^3$
  divergence related to an arbitrary fixing  of the initial fiducial cell. This requires further justification as setting various parameters just to agree with quantum gravity phenomena can simply transfer the issue into why the Planck units take the values they do.  This ``gauge''  factor is instrumental in potentially
  solving the flatness problem when curvature $k\neq 0$ is
  present [26]: various matter components are affected differently by scaling the scale factor. By removing this factor they remove any possibility of
  resolving the flatness problem in the models considered since it
  already exists in the pre-bouncing phase. For example one might
  wonder why a curvature dominated  Milne model (k=-1) does not just perpetually expand
  from time $t=-\infty$ ?

  Here is where a difficulty with setting  up the scalar field arises: if we simply set
  the field at the minimum of the potential it will typically
  oscillate \footnote{We ignore the further issue that one might
   expect  the oscillations about the minimum  to be out of phase beyond some
  initial {\em coherent length scale }, but this would entail using an inhomogeneous metric. Arbitrary small
  initial amplitudes for $\phi$  would contradict quantum uncertainty conditions. }
 about the minimum with a high frequency $m\sim 10^{-5}$ and as the
  universe contracts the amplitude of the field $\phi$ will
  increase as typically $\sim a^{-3/2}$ e.g. [1,2,27]. But now whatever the
  couplings are that typically  reheat  after any inflationary
  phase, where again the field oscillates with frequency $m$, will interfere  and dissipate this high
  frequency oscillation into particle creation as the universe
  slowly contracts: so either the couplings are mysteriously absent and the universe would always be dominated by the
  scalar field and any subsequent inflationary phase could not reheat by the usual
  oscillating scalar field mechanism; or the
  reheating effectively occurs during the final stages of  collapse preventing any large and {\em coherent} scalar field
   being present and   no subsequent inflation caused by $V(\phi)$ would result. This problem has been
  overlooked cf.[3] because ``the reheating has simply been assumed to
  occur after rolling into the minimum but not when it already was in the minimum'' so introducing an
   asymmetry that is unwarranted. There might be convoluted ways
   of escaping this dilemma but at first sight  is seems there
   will be entropy production as the universe collapses due to the high frequency
    oscillations of $\phi$ together with the presence of
   any remotely non-zero couplings to other matter, essentially because the mass $m$ has to set
    the scale of perturbations during
   the inflationary phase but is also responsible for  high frequencies oscillations when $H<m$ : $\sim 10^{-5}$ in Planck
   units is vastly above, for example,  electroweak scales. This
    then suggests the universe will become simply  dominated by
   standard non-inflationary matter as the bounce approaches and
   so does not  allow  any inflationary low entropic conditions
   to naturally occur.

    The scalar field development would therefore be
    more realistically  of the form\\
    \\
   Osc $\rightarrow$ Matter $\stackrel{\times}{\rightarrow}$
   Potential $\rightarrow $ Osc $\rightarrow$ Matter\\
   \\
   with a bottleneck developing before potential domination can
   occur which would apparently require violation of the
   generalized second law to proceed in the required manner. One
   naturally expects entropy to increase from left to right in this
   development and the entropy $S$  of a potential driven inflationary
   phase is comparatively small $S\simeq 1/V(\phi)$ [1,2]. In ref.[25] they obtained a limit, on their variable $F_B$,  which corresponds to entropy $S\sim 1/F_B \leq 10^{10}$ so being very restrictive on the pre-bouncing phase. Even if matter was not formed, because of extremely
   weak coupling,  it is known that
   climbing the potential is unstable e.g.[28] and the scalar field will instead become kinetic energy dominated so the next most realistic scenario would be \\
   \\
   Osc $\rightarrow$ Kinetic energy $\rightarrow$ Osc $\rightarrow$ Matter\\
   \\
   In neither case is it realistic to obtain that the potential would come to dominate and so drive a subsequent stage of inflation.

   In conclusion, we have outlined problems for the bouncing
   scenario either with just standard matter or when a scalar field is
   included to hopefully drive an extra inflationary phase. Either
   large initial
   entropy or growth during the semi-eternal
   collapsing phase, due to
   quantum dissipative effects\footnote{Even if the effects are mainly occurring just seconds before the bounce one generally
    needs reheating fairly rapidly, so the couplings cannot be excessively weak,
     to be compatible with nucleosynthesis constraints [1,2].} , will tend to cause violation of
   the holography bound during any possible bounce - so in apparent contradiction
   with the claims in [29]. Any subsequent useful inflation
   caused by any generated homogeneous scalar field is not compatible with
   placing the field simply at the minimum of a potential with,
   however slight, couplings to standard  matter.

   We finally note that there is some similarity with bouncing
   models occurring in Horava-Lifshitz gravity to those considered here-see
   eg.[30]. It would be of interest to see if they also suffer from
   related concerns.

\newpage
{\bf References}\\
\begin{enumerate}

\item A.D. Linde, ``Particle Physics and Inflationary cosmology''
(Harwood Press)  1990.
\item E.W. Kolb and M.S. Turner, ``The Early Universe''
(Addison-Wesley: New York) 1990.

\item P. Singh, K. Vandersloot and G.V. Vereshchagin, Phys. Rev. D
74 (2006) p. 043510.
\item P. Singh, K. Vandersloot and G.V.
Vereshchagin, Phys. Rev. Lett. 96 (2006) p.141301
\item A. Ashtekar, T. Pawlowski, P. Singh and K. Vandersloot,
Phys. Rev. D 75 (2007) p.024035.
\item A. Ashtekar, A. Corichi and P. Singh, Phys. Rev. D 77 (2008)
p.024046.

\item Y. Shtanov and V. Sahni, Phys. Lett. B 557 (2003) p.1.\\
M.G. Brown, K. Freese and W.H. Kinney, JCAP 3 (2008) p.002.
\item G. Dvali, G. Gabadadze and G. Senjanovic, hep-ph/9910207.\\
{\em see also}  I. Quiros,  arXiv:0706.2400

\item L. Baum and P.H. Frampton, Phys. Rev. Lett. 98 (2007)
p.071301.
\item C. Barcelo and M. Visser, Phys. Lett. B 482 (2000) p.183.
\item G.'t Hooft,  gr-qc/9310026.\\
L. Susskind, J. Math. Phys. 36 (1995) p.6377.
\item D. Bigatti and L. Susskind, hep-th/0002044.\\
R. Bousso, Rev. Mod. Phys. 74 (2002) p.825.
\item J.D. Barrow, New. Astron. 4 (1999) p.333.
\item J.D. Barrow and M.P. Dabrowski, Mon. Not. R. Astron. Soc.
275 (1995) p.850.
\item R. Durrer and J. Laukenmann, Class. Quant. Grav. 13 (1996)
p.1069.
\item L. Parisi, M. Bruni, R. Maartens and K. Vandersloot, Class.
Quant. Grav. Grav. 24 (2007) p.6243.
\item P.J. Steinhardt and N. Turok, Phys. Rev. D 65 (2002)
p.126003.
\item J.K. Erickson, D.H. Wesley, P.J. Steinhardt and N. Turok,
Phys. Rev. D 69 (2004) p.063514.
\item A. Corichi, T. Vukasinac and J.A. Zapata, Phys. Rev. D 76
(2007) p.0440163.
\item A. Ashtekar, J. Baez, A. Corichi and K. Krasnov, Phys. Rev.
Lett. 80 (1998) p.904.
\item R.D. Sorkin, R.M. Wald and Z.J. Zhang, Gen. Rel. Grav. 13
(1981) p.1127.\\
S.D.H. Hsu and D. Reeb, Phys. Lett. B 658 (2008) p.244.\\
{\em see also} P.H. Frampton, S.D.H. Hsu, D. Reeb and T.W.
Kephart, Class. Quant. Grav. 26 (2009) p.145005.
\item D.H. Coule, Class. Quant. Grav. 22 (2005) p.R125.\\
D.H. Coule, arXiv:0706.0205. [gr-qc].
\item M. Bojowald, Nature 3 (2007) p.523.\\
A. Corichi and P. Singh, Phys. Rev. Lett. 100 (2008) p.161302.
\item G.W. Gibbons, S.W. Hawking and J.M. Stewart, Nucl. Phys. B
281 (1987) p.736.\\
M. Henneaux, Nuovo Cimento Lett. 38 (1983) p.609.
\item A. Ashtekar and D. Sloan, Phys. Lett. B 694 (2010) p.108.\\
A. Corich and A. Karami, Phys. Rev. D 83 (2011) p.104006.\\
A. Ashtekar and D. Sloan, Gen. Rel. Grav. 43 (2011) p.3619.
\item S.W. Hawking and D.N. Page, Nucl. Phys. B 298 (1988)
p.789.\\
D.H. Coule, Class. Quant. Grav. 12 (1995) p.455.
\item M. Giovannini, ``A primer on the physics of the cosmic
microwave background'', (World Scientific: Singapore) 2008
\item D.S. Goldwirth, Phys. Lett. B 256 (1991) p.354.
\item A. Ashtekar and E. Wilson-Ewing, Phys. Rev. D 78 (2008)
p.06407.\\
 A. Ashtekar, Gen. Rel. Grav. 41 (2009) p.707.
\item R. Brandenberger, Phys. Rev. D 80 (2009) p.043516.
\end{enumerate}
\end{document}